\documentclass[fleqn,usenatbib,referee]{mnras}

\usepackage{newtxtext,newtxmath}

\usepackage[T1]{fontenc}

\DeclareRobustCommand{\VAN}[3]{#2}
\let\VANthebibliography\thebibliography
\def\thebibliography{\DeclareRobustCommand{\VAN}[3]{##3}\VANthebibliography}

\usepackage{graphicx}	
\usepackage{amsmath}	
\usepackage{xcolor}

\title[Swift J0243.6+6124]{Timing analysis of Swift J0243.6+6124 with \emph{NICER} and \emph{Fermi}/GBM during the decay phase of the 2017--2018 outburst}

\author[M. M. Serim et al.]{
M. M. Serim$^{1}$\thanks{E-mail: serim@astro.uni-tuebingen.de},
Ç. K. Dönmez$^{2}$,
D. Serim$^{1}$, L. Ducci$^{1}$ ,  A. Baykal$^{2}$, and A. Santangelo$^{1}$
\\
$^{1}$Institut fur Astronomie und Astrophysik, Eberhard Karls Universit\"{a}t, Sand 1, D-72076 Tübingen, Germany\\
$^{2}$Department of Physics, Middle East Technical University, 06800 Ankara, Turkey\\
}

\date{Accepted XXX. Received YYY; in original form ZZZ}

\pubyear{2023}

\begin{document}
\label{firstpage}
\pagerange{\pageref{firstpage}--\pageref{lastpage}}
\maketitle

\begin{abstract}
We present a timing and noise analysis of the Be/X-ray binary system Swift J0243.6+6124 during its 2017--2018 super-Eddington outburst using \emph{NICER}/XTI observations. We apply a synthetic pulse timing analysis to enrich the \emph{Fermi}/GBM spin frequency history of the source with the new measurements from \emph{NICER}/XTI. We show that the pulse profiles switch from double-peaked to single-peaked when the X-ray luminosity drops below $\sim$$7\times 10^{36}$ erg s$^{-1}$. We suggest that this transitional luminosity is associated with the transition from a pencil beam pattern to a hybrid beam pattern when the Coulomb interactions become ineffective to decelerate the accretion flow, which implies a dipolar magnetic field strength of $\sim$$5\times 10^{12}$ G. We also obtained the power density spectra (PDS) of the spin frequency derivative fluctuations. The red noise component of the PDS is found to be steeper ($\omega^{-3.36}$) than the other transient accreting sources.
We find significantly high noise strength estimates above the super-Eddington luminosity levels, which may arise from the torque fluctuations due to interactions with the quadrupole fields at such levels.
\end{abstract}




\section{Introduction}
A new transient binary system in our galaxy, Swift J0243.6+6124, was discovered at the onset of the outburst phase \citep{2017Kennea}.
Initial analyses showed that the system consists of a pulsar with a $\sim$9.8 s spinning period \citep{2017Kennea} and an O9.5Ve type companion with long-term optical and infrared variabilities similar to the common Be/X-ray binary systems \citep{2017Kouroubatzakis, 2020Reig}.
Using optical observations, \cite{2020Reig} estimated the distance of the system to be $\sim$5 kpc, whereas the \emph{Gaia} EDR2 estimated distance was 6.8$_{-1.1}^{+1.5}$ kpc \citep{2018BailerJones}.
Adopting the \emph{Gaia} EDR2 distance, the maximum brightness of Swift J0243.6+6124 was estimated to be $\sim$2 $\times$ 10$^{39}$ erg s$^{-1}$ at the peak of the outburst \citep{2018Tsygankov, 2020Doroshenko}. On the other hand, the source distance (id: 465628193526364416) is revised as $5.2\pm0.3$ kpc in the \emph{Gaia} EDR3 catalogue \citep{2021Bailer}.
When the distance of $\sim$5 kpc is taken into account, the peak luminosity would be $\sim$1 $\times$ 10$^{39}$ erg s$^{-1}$, which is still higher than the Eddington limit for such a neutron star \citep{2020Reig}; thus, Swift J0243.6+6124 is classified as an ultraluminous X-ray Pulsar (ULXP), the first ever detected in our own galaxy.

Numerous studies regarding the temporal and spectral properties of Swift J0243.6+6124 have been conducted in attempt to comprehend the physical dynamics of this unique source \citep{2018WilsonHodge, 2018Tsygankov, 2018VanDenEijnden, 2018Jaisawal, 2019Tao, 2020Doroshenko, 2020Sugizaki, 2020Kong, 2020Wang, 2022Kong, 2022Liu, 2022Bykov}.
Detailed investigations revealed that both temporal and spectral features, including shape of the power spectra, pulse profiles and energy spectra, change systematically at two different transitional luminosity levels (for 6.8 kpc), $L_1 \sim 1.5 \times 10^{37}$ erg s$^{-1}$ and $L_2 \sim 4.5 \times 10^{38}$ erg s$^{-1}$ \citep{2018WilsonHodge, 2020Doroshenko, 2020Kong}.
Thus, these transitional luminosity levels $L_1$ and $L_2$ are interpreted as transitions of subcritical to supercritical accretion regime and supercritical to radiation pressure dominated disc (RPD) accretion regime, respectively \citep{2020Doroshenko}.
As a reminder for the discussions throughout the paper, using the new Gaia distance of 5.2 kpc, the transitional luminosity levels are calculated as $L_1$ = $8.8 \times 10^{36}$ erg s$^{-1}$ and $L_2$ = $2.6 \times 10^{38}$ erg s$^{-1}$, respectively.

Despite the extensive studies, the magnetic field configuration of Swift J0243.6+6124 is not yet clear.
Initial studies have demonstrated that the source pulsations are still detectable at luminosities as low as 10$^{34}-10^{35}$ erg s$^{-1}$, which indicates that the propeller regime has not yet been attained at such low luminosities; consequently, the pulsar should have a very compact magnetosphere to allow accretion to continue, which confines the upper limit of the magnetic field strength to 3 $\times$ 10$^{12}$ G \citep{2018Tsygankov, 2020Doroshenko}.
Phase-resolved spectral analysis of \emph{NuSTAR} observations at different luminosity levels hints for a thick super-Eddington disc with an inner radius of 2--3 $\times$ 10$^{7}$ cm and a weakly variable reflection component, signifying a magnetic field strength $3\times10^{12}$ G if the field is dipolar \citep{2022Bykov}.
On the other hand, the discovery of a cyclotron resonance scattering feature (CRSF) in the spectrum of Swift J0243.6+6124 at $\sim$120--146 keV, which is only visible in certain phases around the peak of the outburst \citep{2022Kong} implies a magnetic field strength of $\sim$1.6 $\times$ 10$^{13}$ G near the surface of the pulsar.
Nevertheless, it is suggested that the observed CRSF is actually associated with multipole fields \citep{2022Kong} and the dipolar component of the field strength should be in the range of 3--9 $\times$ 10$^{12}$ G in order to describe the observed properties of the source coherently \citep{2020Doroshenko}.
The accretion disc possibly penetrates into the magnetosphere more than expected, and the disc interactions are dominated by multipole components of the field at high luminosities \citep{2020Doroshenko, 2022Kong}.

With its ultraluminous episode and unique properties, the source has been the target of many studies, especially in probing the nature of neutron star accretion at very high luminosities \citep{2018VanDenEijnden, 2018Doroshenko, 2018WilsonHodge, 2019Jaisawal, 2020Kong, 2022Kong, 2022Bykov}.
In this study, we investigate the timing properties of Swift J0243.6+6124, focusing mostly on its moderately luminous stages ($\sim$10$^{36}-10^{37}$ erg s$^{-1}$) towards the end of the outburst in 2017--2018, during which the source remained in a subcritical accretion state.
We describe the data and the relevant screening processes used for timing analysis in Section \ref{sec2}.
In Section \ref{sec3}, we represent the pulse timing analyses that are used for measuring spin frequencies and generating pulse profiles. In addition, we also demonstrate our results on the torque fluctuations on different timescales and luminosities. 
Lastly, in Section \ref{sec4}, we review and discuss the results of our study in the light of the systematic luminosity-dependent evolution of pulse profiles.

\section{Data}
\label{sec2}
Neutron Star Interior Composition Explorer (\emph{NICER}) is stationed on International Space Station (ISS) since 2017 June and operated by NASA.
Its primary instrument, X-Ray Timing Instrument (XTI), consists of an aligned array of 56 X-ray concentrators and focal plane modules (FPM) collecting photons from a $\sim$30 arcmin$^2$ field onto silicon field detectors in each FPM.
These detectors are capable of soft X-ray spectroscopy with 0.2--12 keV energy range and <300 ns timing precision with $\sim$1900 cm$^2$ cumulative effective area at 1.5 keV \citep{2016Gendreau}.

Swift J0243.6+6124 has 214 \emph{NICER}/XTI observations in the \emph{NICER} master catalogue between 2017 October 03 and 2019 June 07 (MJD 58029--58641), corresponding to the outbursts in this study.
Among those, we utilise the observations prior to the rapid decline of the source luminosity at the end of 2019 February.
Data reduction of the observations is done with \textsc{heasoft} v6.29 using the most recent calibration files at the time (CALDB release \texttt{xti20210707}) for \emph{NICER}.
The clean events and filter files for screening data are reproduced by employing the standard level 2 data processing steps provided by the \texttt{nicerl2} tool.
Good time intervals (GTI) are selected adopting the default screening parameters recommended by the NICER team\footnote{\url{https://heasarc.gsfc.nasa.gov/lheasoft/ftools/headas/nicerl2.html}}: ISS is outside the predefined Southern Atlantic Anomaly (SAA) region, a minimum of 38 of the 56 detectors are enabled, the pointing offset is less than 0.015$^{\circ}$, the source is at least 15$^{\circ}$ away from the dark Earth limb and 30$^{\circ}$ away from the bright Earth limb.
The event files between MJD 58029--58531 are merged using the \texttt{nimpumerge} tool, event time-series are barycentered using \texttt{barycorr} with the JPL ephemeris DE430, and the light curves with a time resolution of 0.1 s are extracted with \texttt{XSELECT}.
We also corrected the photon arrival times of the generated \emph{NICER}/XTI light curve prior to the timing analyses described below with the orbital solution provided by the \emph{Fermi}/GBM Accreting Pulsars Program (APP) team (See Table \ref{table:orbit}).
\begin{table}
	\centering
	\caption{Orbital parameters provided by the \emph{Fermi}/GBM team, which are used to correct the photon arrival times prior to our timing analysis.}
		\begin{tabular}{l | c | c}
        \hline\hline
        $P_{\textrm{orb}}$ & 27.698899 & days \\
		$T_{\pi/2}$ & 58116.097 & MJD \\
		$a_x$ sin $i$ & 115.531 & lt-s \\
		$\omega$ & -74.05 & degrees \\
        $e$ & 0.1029 & -- \\
        \hline
		\end{tabular}
	\label{table:orbit}
\end{table}

Additionally, we make use of the pulse frequency history of Swift J0243.6+6124 which is publicly shared through \emph{Fermi}/GBM (Gamma-ray Burst Monitor) monitoring program website\footnote{\label{note1}\url{https://gammaray.nsstc.nasa.gov/gbm/science/pulsars/lightcurves/swiftj0243.html}} \citep{2020Malacaria} and the regularly updated \emph{Swift}/BAT light curves maintained by the \emph{Swift}/BAT team\footnote{\url{https://swift.gsfc.nasa.gov/results/transients/weak/SwiftJ0243.6p6124}} \citep{2013Krimm}.
The compiled \emph{Fermi}/GBM frequency history is orbit-corrected and encapsulates the range between 2017 October 01 and 2019 January 14 (MJD 58027--58497).
We used the \emph{Swift}/BAT daily average light curve, which has an energy range of 15--50 keV, from the discovery of the source in 2017 October up to 2019 February.
\citet{2020Doroshenko} argued that 2--150 keV count rates for \emph{Insight-HXMT} appear to be consistent with those measured by \emph{Swift}/BAT, and the \emph{Swift}/BAT count rates can be roughly converted to bolometric luminosity using a scaling factor $\sim$8.2 $\times$ 10$^{38}$, assuming a source distance of 6.8 kpc. In this article, we utilise the \emph{Gaia} EDR3 distance (5.2 kpc) and revise scaling factor for the \emph{Swift}/BAT count rate--luminosity conversion to $\sim$4.8 $\times$ 10$^{38}$ to estimate the bolometric luminosity, unless otherwise stated.

\section{Timing Analysis and Results}
\label{sec3}
\subsection{Synthetic pulse timing}
\label{sec31}
During its outburst phase in 2017--2018, the X-ray luminosity of Swift J0243.6+6124 varies by five orders of magnitude.
At the same time, the accretion geometry, and consequently the pulse profiles, drastically alter at different accretion regimes \citep{2018WilsonHodge, 2020Doroshenko}.
In particular, the pulse profiles are shown to be double-peaked at subcritical regime ($L_x < L_1$) and evolve into a single-peaked shape at supercritical regime ($L_1 < L_x < L_2$), then again transform into a double-peaked structure at the highest luminosities ($L_2 < L_x$) \citep{2020Doroshenko}.
Moreover, the spin-up rate during the initial stages of the outburst is very strong, reaching up to 2.2 $\times$ 10$^{-10}$ Hz s$^{-1}$ \citep{2018Doroshenko}.
The frequency derivative leads to a phase shift of one cycle on a timescale of $\sqrt{2/|\dot{\nu}|}$ \citep{2014Acuner}; and with the reported high spin-up rate during the outburst \citep{2018WilsonHodge, 2018Doroshenko}, this timescale becomes as short as $\sim$1.1 days.
Combined with the pulse profile variations, employing phase-coherent timing technique becomes unfavourable at the luminous stages of the outburst.
\begin{figure}
\centering
	\includegraphics[width=0.9\columnwidth]{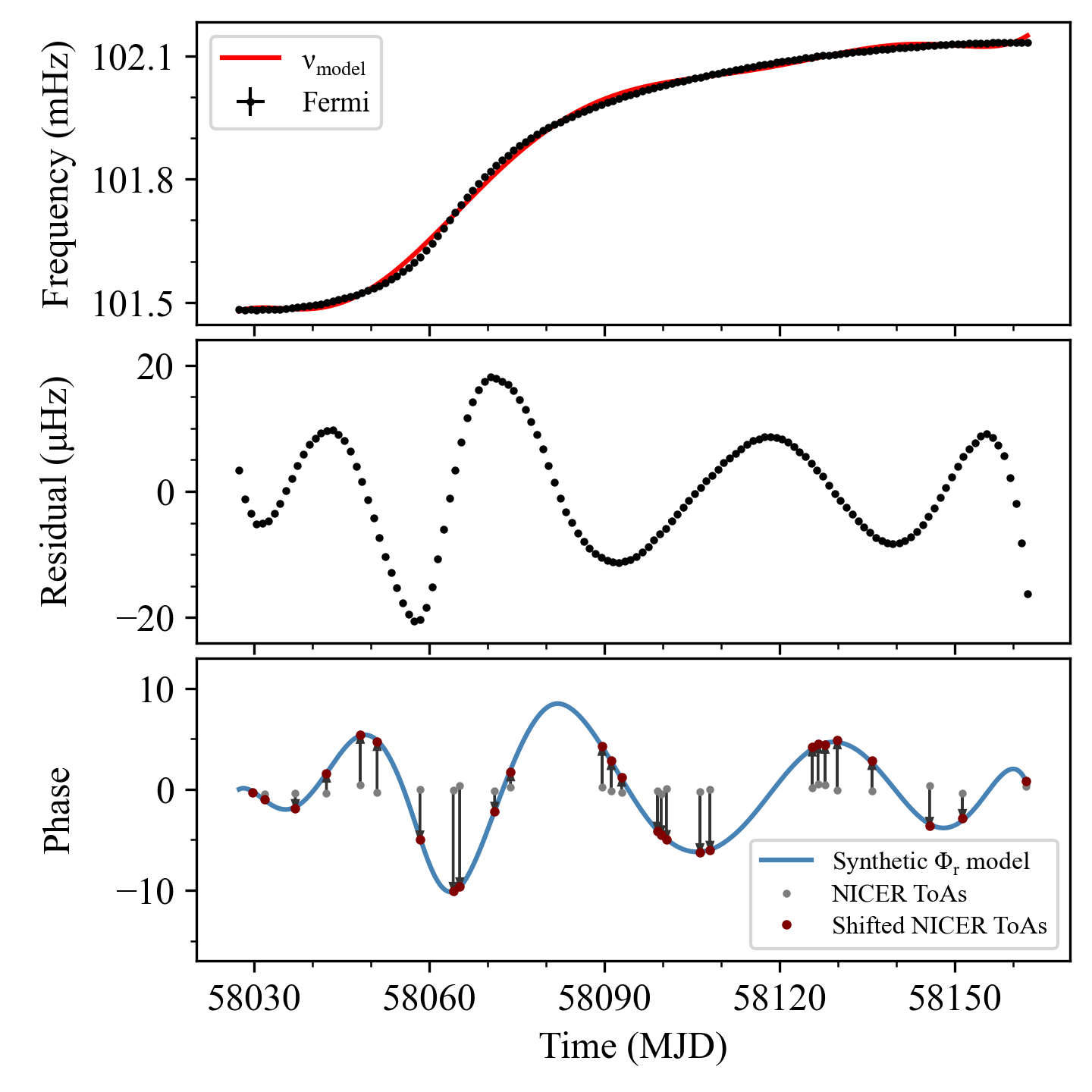}
    \caption{Upper panel: \emph{Fermi}/GBM spin frequency history (black dots) and the synthetic timing solution (red curve) obtained from polynomial fitting in the interval 1 (see Table \ref{table:timing} for timing parameters). Middle panel: The frequency residuals of the synthetic timing solution. Lower panel: The synthetic phase residual model obtained from \emph{Fermi}/GBM residuals (blue curve), the residuals of time of arrivals (TOAs) of the \emph{NICER} observations (gray dots) when folded with the corresponding timing solution and the shifted \emph{NICER} TOA residuals (red dots) according to the expected phase residual model.}
    \label{fig:interval1}
\end{figure}
Making use of the refined orbital solution provided by \emph{Fermi}/GBM team, we used the following approach to measure the pulse frequencies from the \emph{NICER} data, which reside within the same time interval as the \emph{Fermi}/GBM measurements:
We first divide the \emph{Fermi}/GBM pulse frequency measurements into three different segments, each of which is fitted with a different polynomial model to represent the frequency evolution over time, and obtain a synthetic timing solution\footnote{At this stage, it should be noted that the choice of a polynomial model order is rather arbitrary; nonetheless, the synthetic residual reconstruction compensates for the possible deviations from the model. In principle, the procedure can be applied for any polynomial order, provided that the reconstructed pulse profiles are compatible with the actual profiles.} (see Table \ref{table:timing}).
Using these timing solutions, we then calculate the deviations of the \emph{Fermi}/GBM frequencies from the model to extract its residuals.
Utilising a linear spline interpolation of the \emph{Fermi}/GBM frequency residual data set $\nu_{r}$, we convert them to a synthetic phase residual model $\Phi_{r}$ using integration:
\begin{equation}
\Phi_{r}(t) = \int_{t_0}^{t} \nu_{r}(t')dt'
\end{equation}
where $t_0$ indicates the start time of the segment.
Next, we fold the orbitally-corrected \emph{NICER} light curve with the same synthetic timing solution to generate its phase residuals.
Finally, we shift the \emph{NICER} phase residuals to match with the synthetic phase residual model obtained from \emph{Fermi}/GBM (see Figure \ref{fig:interval1}).

In the first interval, the luminosity of the source changes substantially, resulting in significant deviations from the polynomial description of the rapid frequency evolution.
However, the phase residuals of \emph{NICER} observations become compatible with the synthetic residual model when they are shifted with expected integers in phase domain.
The only exceptions are the pulse profile variations at different episodes \citep{2020Doroshenko} that are needed to be taken into account.
Thus, we further allow phase shifts for the pulses in the supercritical regime by $\Delta\phi \sim 0.5$, corresponding to the phase difference between the peaks of the double-peaked and one-peaked profiles (see Figure 4 of \citet{2020Doroshenko}) to accord them with the expected synthetic phase residuals (See Figure \ref{fig:interval1}, bottom panel).
On the other hand, during the late stages of the outburst (at the interval 2, 3 and 4), the source luminosity is rather low ($L_x\lesssim8\times10^{37}$ erg s$^{-1}$), and Swift J0243.6+6124 continues to accrete only in subcritical regime (i.e. $L_x<L_1$).
Therefore, the synthetic phase residuals reside within a single cycle for the corresponding synthetic timing solutions given in Table \ref{table:timing}.

\begin{figure*}
	\includegraphics[width=\columnwidth]{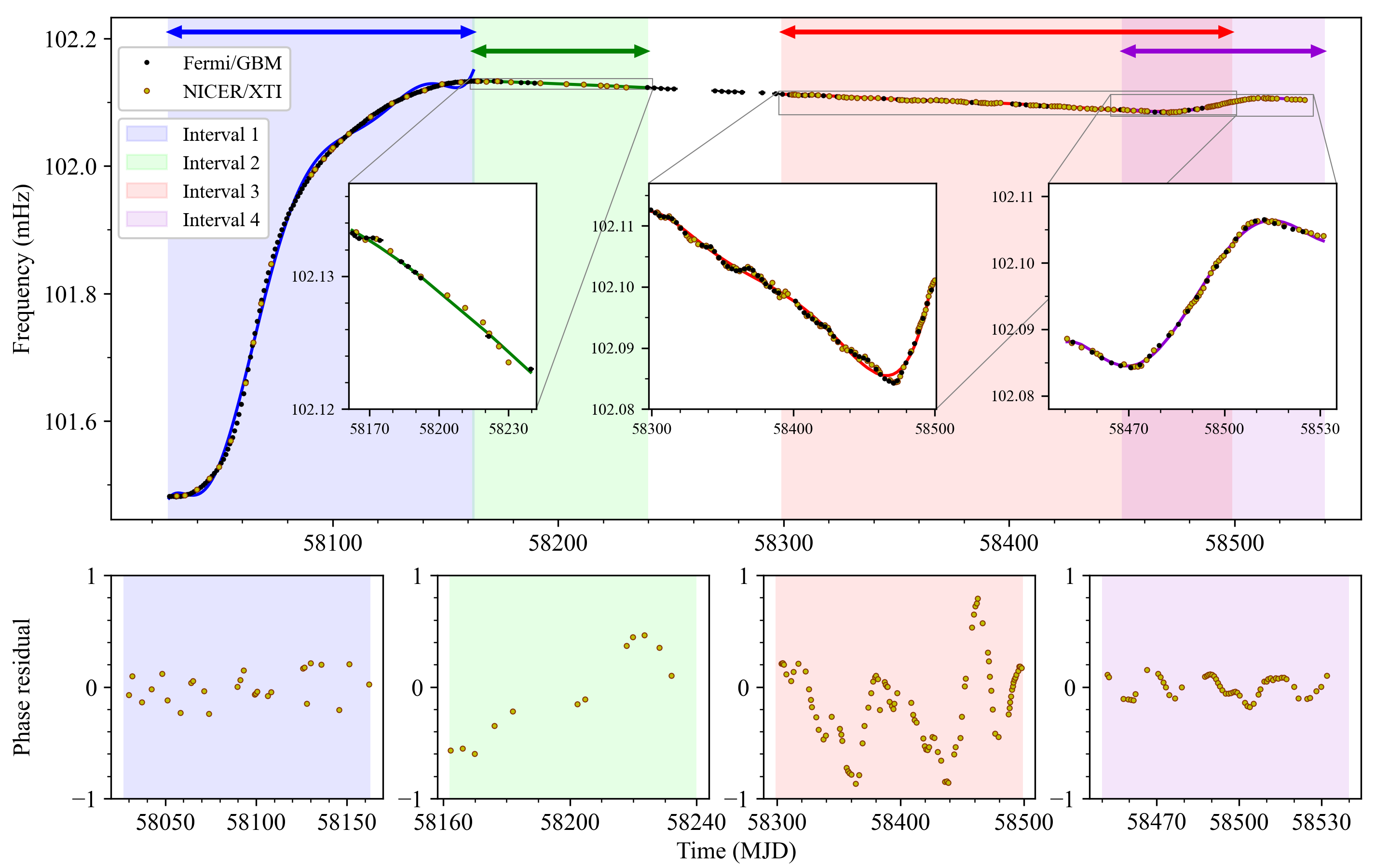}
    \caption{Top panel: The spin frequency history of Swift J0243.6+6124 generated for all intervals. The shaded regions represent corresponding time intervals. The inset panels show the same frequencies in the marked regions with a different scale for better viewing. Bottom panels: The resultant \emph{NICER} TOA residuals of the synthetic timing solutions.}
    \label{fig:allintervals}
\end{figure*}

Finally, in order to convert synthetic timing solutions to pulse frequency measurements, we use each consecutive pair of pulse arrivals in the \emph{NICER} residual set.
Each pair is fitted with a linear function $\delta \nu = \delta \phi /(t_2 - t_1)$ where $t_1$ and $t_2$ are the arrival times of the first and second pulses in the pair, and $\delta\phi$ is the phase difference between the pair used for fitting.
Each fit is transformed into a spin frequency measurement at the corresponding interval's midpoint by using the frequency correction $\delta\nu$ over the synthetic timing solutions (for applications, see \citet{2019CerriSerim, 2021Serim}).
The 1$\sigma$ error ranges of the slope are used as a gauge of the uncertainty in the spin frequency measurements.
Figure \ref{fig:allintervals} demonstrates the spin frequency history measured from \emph{NICER} observations whose results are consistent with the spin frequency history shared by the \emph{Fermi}/GBM APP team.

As it can be seen from the frequency history presented in Figure \ref{fig:allintervals}, apart from the initial stages of the Type II outburst, Swift J0243.6+6124 also spins up between MJD $\sim$58470--58510. Afterwards, as the flux diminishes over time, the frequency evolution trend returns back to the spin-down stage with an average frequency derivative of $\sim-1.6 \times 10^{-12}$ Hz s$^{-1}$, which is comparable to the average spin-down rate $\sim -1.8 \times 10^{-12}$ Hz s$^{-1}$ observed between MJD 58150--58460.

\begin{table*}
	\centering
	\caption{The timing parameters obtained for four different intervals. Note that the synthetic timing solutions for all the intervals are established by fitting the existing \emph{Fermi}/GBM frequencies.}
	\small{
		\begin{tabular}{l c c c c}
                                                     & Interval 1     & Interval 2    & Interval 3    & Interval 4 \\ \hline
			\vspace{1mm}
			Timing range (MJD)                       & 58027--58162   & 58162--58239  & 58299--58497  & 58450--58531 \\
			\vspace{1mm}
			Epoch (MJD)                              & 58050          & 58200         & 58350         & 58500 \\
			\vspace{1mm}
			$\nu$ (Hz)                               & 0.1015335(20)  & 0.1021289(2)  & 0.1021046(2)  & 0.10210146(7)  \\
			\vspace{1mm}
			d$\nu$/dt (10$^{-10}$ Hz s$^{-1}$)       & 1.02(2)        & $-$0.0161(4)  & $-$0.0180(9) & 0.0730(10) \\
			\vspace{1mm}
			d$^2\nu$/dt$^2$ (10$^{-17}$ Hz s$^{-2}$) & 9.85(31)       & $-$0.0109(45) & 0.0265(36)    & $-$0.408(13) \\
			\vspace{1mm}
			d$^3\nu$/dt$^3$ (10$^{-24}$ Hz s$^{-3}$) & $-$49.4(35)    & -             & $-$0.0158(32) & $-$4.52(23) \\
			\vspace{1mm}
			d$^4\nu$/dt$^4$ (10$^{-31}$ Hz s$^{-4}$) & $-$9.86(90)    & -             & $-$0.0139(22) & 35.9(29) \\
			\vspace{1mm}
			d$^5\nu$/dt$^5$ (10$^{-36}$ Hz s$^{-5}$) & 158.2(109)     & -             & 0.0556(51)    & 3.69(41) \\
			\vspace{1mm}
			d$^6\nu$/dt$^6$ (10$^{-41}$ Hz s$^{-6}$) & $-$9.77(66)    & -             & -             & $-$0.114(59) \\
			\vspace{1mm}
			d$^7\nu$/dt$^7$ (10$^{-47}$ Hz s$^{-7}$) & 2.41(17)       & -             & -             & - \\ \hline
		\end{tabular}}
	\label{table:timing}
\end{table*}

\begin{figure}
\centering
	\includegraphics[width=0.8\columnwidth]{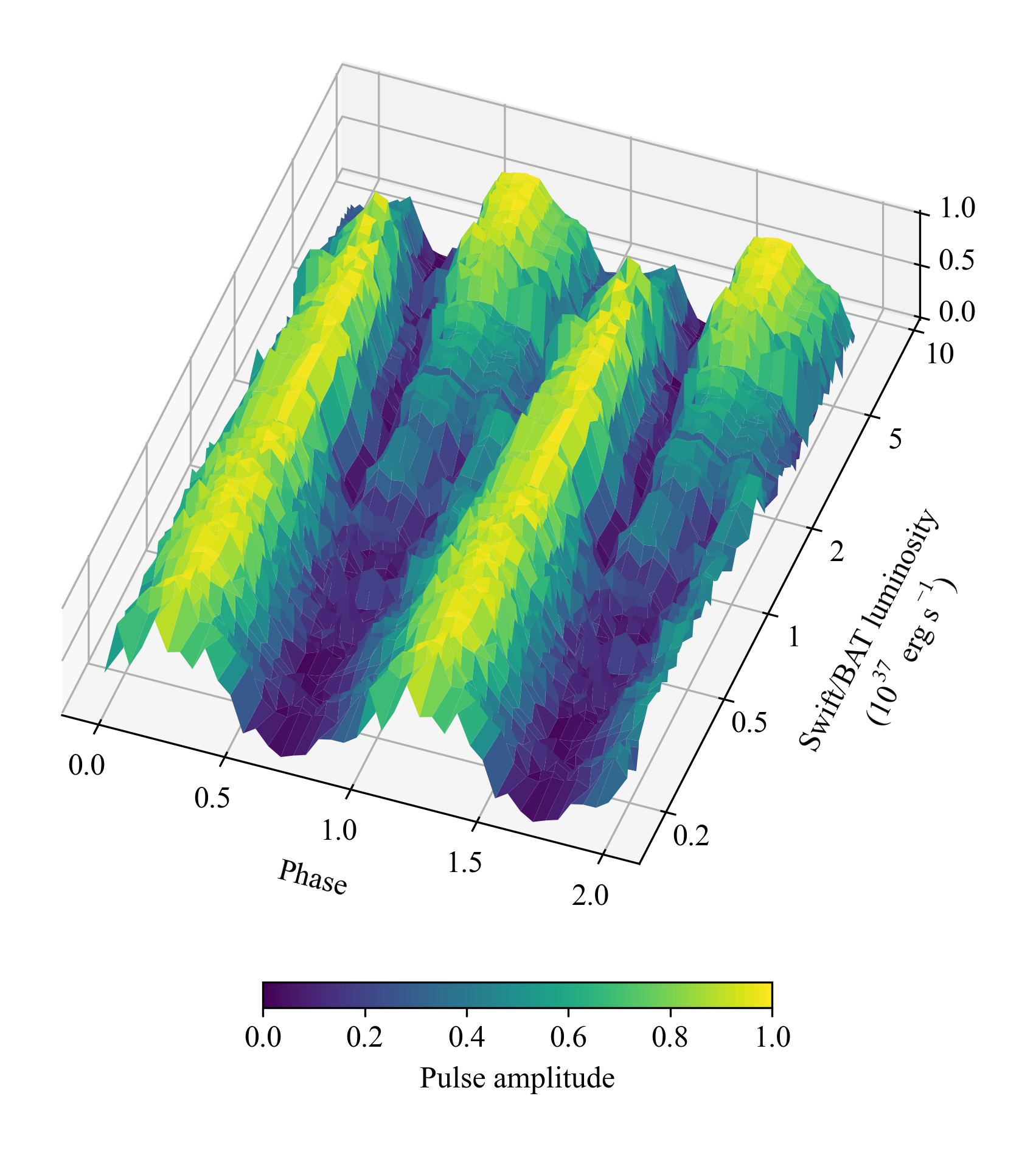}    
    \caption{Luminosity-dependent pulse profile evolution of Swift J0243.6+6124 for the observations after MJD 58300. The pulse profiles are normalized to [0, 1] and plotted for two cycles for clarity.}
    \label{fig:pulseprofiles}
\end{figure}

Interestingly, when the pulse profiles obtained from the timing analysis at low flux states are examined, the pulses seem to exhibit single-peaked profiles at very low flux levels. To illustrate this behaviour more clearly, we present the luminosity-sorted pulse profiles (normalized to [0, 1] range) of all observations after MJD 58300 in Figure \ref{fig:pulseprofiles}. A systematic change in the profiles emerges at a luminosity level of $\sim$$7\times10^{36}$ erg s$^{-1}$, marking a potential new transitional level for the alteration of the accretion geometry. As the luminosity decreases, the main peak gradually fades away and the secondary peak grows stronger. It is also interesting to note that the source tends to exhibit spin-down episodes below this luminosity level.

\subsection{Timing noise}
Using the whole frequency history enriched with \emph{NICER} measurements, we investigate the temporal noise behaviour of Swift J0243.6+6124.
To estimate the amplitude of the timing noise at different timescales, we proceed with the rms-value technique developed by \citet{1972Boynton, 1984Deeter, 1985Cordes}.
This technique utilises the rms values of the timing residuals $\langle\sigma_r (m,T)\rangle$ that are acquired after eliminating the polynomial trend of order $m$ from the data set of duration $T$.
Then, the associated noise strength $S_r$ can be calculated via:
\begin{equation}
    S_r =\frac{\langle\sigma_r (m,T)\rangle}{\langle\sigma_r (m,1)\rangle_u}\frac{1}{T^{2r-1}}
\end{equation}
where $r$ specifies the red noise order, and $\langle\sigma_r (m,1)\rangle_u$ denotes the unit noise strength normalization factor for $T=1$ d and $S_r=1$.
Our calculations are performed with the associated normalization factors gauged through direct evaluations \citep[Table 1]{1984Deeter}.
We start by estimating the noise strength of the maximal time span of the data $T_{max}$ and iterate the calculations for halved timescales ($T_{max}/2^n$ for $n=1,2,3,..$).
Then, we aggregate the noise strength measurements in each timescale into logarithmically-binned power density estimates. The uncertainties of power density estimates are determined from 1$\sigma$ confidence intervals, depending on the number of independent noise strength estimates enclosed during generation of the power density estimate in each timescale, as described in \cite{1987Deeter}.
The distribution of the power density estimates as a function of timescale (or analysis frequency $\omega \equiv 1/T$) generates a power density spectrum (PDS) of the spin frequency derivative fluctuations $P_{\Delta\dot{\nu}}$.
We check for the stability of the PDS profile for different polynomial orders and proceed with the lowest stable order, $m=2$, quadratic polynomial trends, for the input frequency series to characterize the regular spin evolution of Swift J0243.6+6124, assuming the residues after the removal of the trend constitute the timing noise.
In order to check the validity of the power density estimates in each time scale, we also present the corresponding measuremental noise levels by taking the measuremental uncertainties of the frequency data set ($\sigma_i$) into account (green crosses in Figure \ref{fig:noise}), which are calculated via $\frac{\sum_i^N\sigma_i^2}{N\,T\langle\sigma_r (m,1)\rangle_u}$. The measuremental noise level provides a precursor to a noise level at which the measuremental error range becomes dominant over the fluctuations in the data set.

\begin{figure}
\centering
	\includegraphics[width=0.9\columnwidth]{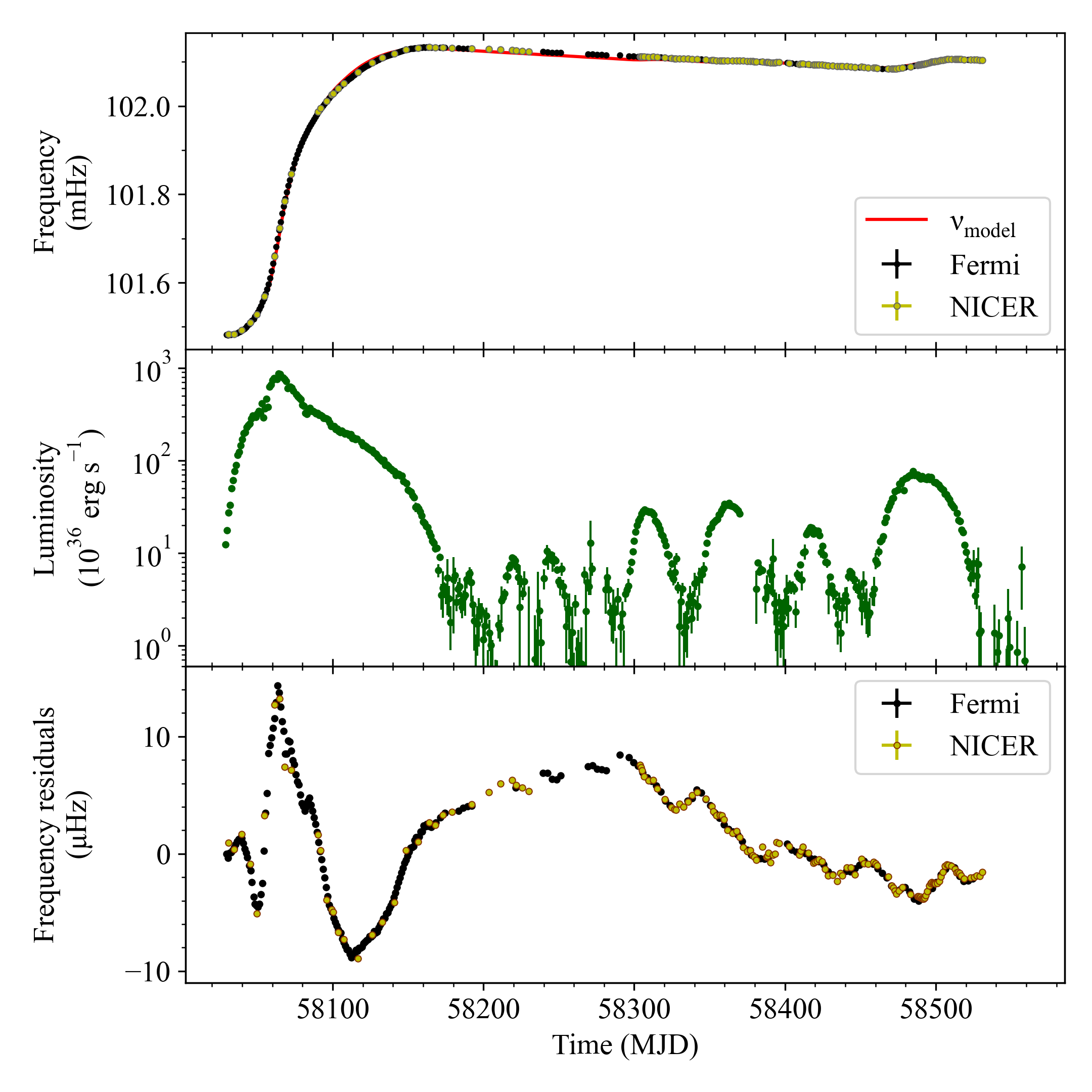}
    \caption{Upper panel: The spin frequency history of Swift J0243.6+6124. Middle panel: Bolometric source luminosity converted from 15--50 keV \emph{Swift/BAT} count rates, assuming $d=5.2$ kpc. Bottom panel: Frequency residuals after eliminating $\nu_{\textrm{model}} (t)$.}
    \label{fig:torquelumfreq}
\end{figure}

In addition to the aforementioned standard method for PDS generation for torque fluctuations, we also follow the approach described in \cite{2021Serim} to see the effects of the accretion torques on the PDS.
In principle, this approach offers a different perspective on the same PDS, with the only distinction being the minimization of torque fluctions arising from disk accretion. It should be noted that in both cases, the input frequency data set is already decoupled from orbital Doppler delays using the orbital parameters given in Table \ref{table:orbit}. Therefore, we assume that the orbital modulations in the frequencies are completely removed and they no longer contribute to the noise strength measurements.
In this case, we utilise a simple power law relation between the spin-up rate and the luminosity, which is modified with a constant spin-down rate ($\dot{\nu}_{model} = \beta L^{\alpha}+ \dot{\nu}_0$) to account for the stable spin-down episodes observed in the frequency history of the source.
Then, the luminosity-dependent frequency evolution model is built as \citep{2021Serim}:
\begin{equation}
    \nu_{model} (t) = \nu_0 + \int_{t_0}^{t} \dot{\nu}_{model} (t') dt'
\end{equation}
where $\nu_0$ is the spin frequency at the time of the burst onset $t_0$.
Instead of polynomial driven residuals built in the standard approach, the residuals obtained from the elimination of $\nu_{model}$ is assumed to inherit the noise component for this case (see Figure \ref{fig:torquelumfreq}).

\begin{figure}
	\includegraphics[width=\columnwidth]{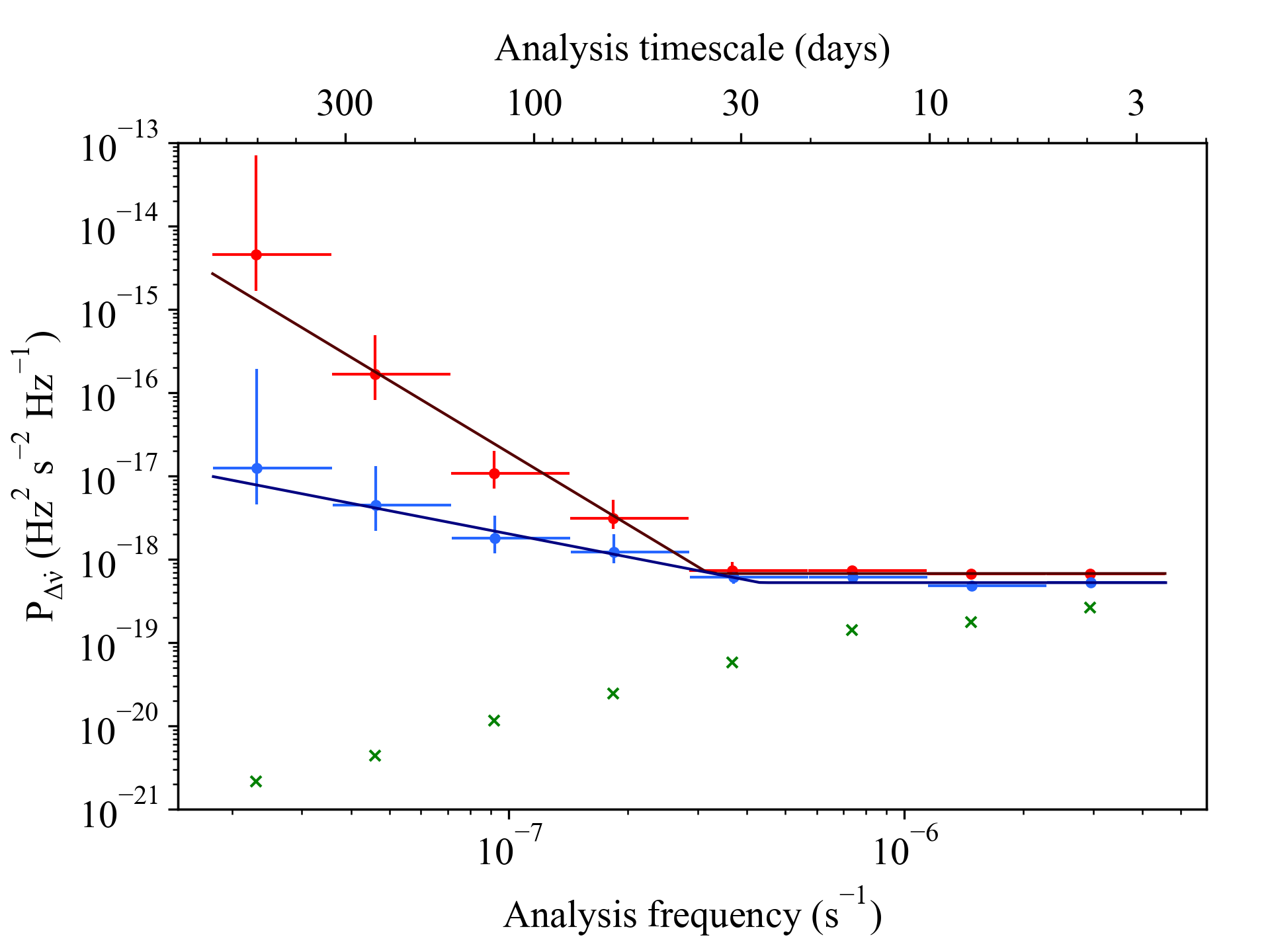}
    \caption{PDS of the spin frequency derivatives using quadratic polynomial trends (red) and luminosity-dependent intrinsic spin frequency evolution model (blue), along with the measuremental noise levels (green). The uncertainties of power density estimates are expressed as 1$\sigma$ confidence intervals, determined by the number of independent estimates present within. Corresponding fits of the PDSs are shown as maroon and dark blue lines.}
    \label{fig:noise}
\end{figure}

In both cases, generated PDSs of spin frequency derivative fluctuations are modeled with a broken power law model:
\begin{equation}
    P_{\Delta\dot{\nu}} =    \begin{cases}
    \, S_{r,1} \, \omega^{\Gamma} \hspace{1cm} \textrm{if}\, \omega<\omega_{\textrm{b}}\\
    \, S_{r,2} \hspace{1.5cm} \textrm{if}\, \omega>\omega_{\textrm{b}}\\
                            \end{cases}
\end{equation}
where $\omega_{b}$ is the break analysis frequency and $\Gamma$ is the power law index of the red noise component
(Figure \ref{fig:noise}). The fitting procedure is carried out by orthogonal distance regression (ODR) using the Python library of \textit{SciPy}. We report the uncertainties of the best fit parameters with $1\sigma$ confidence level.

For the standard approach, in which polynomial trends are used, we find that the PDS of the frequency derivatives is evolving as $\omega^{-3.36\pm0.64}$ within the range $1/46\gtrsim\omega\gtrsim 1/500$ days$^{-1}$, which points out a steeper red noise component when compared with the other accreting sources \citep{1997Bildsten, 2007Baykal, 2021Serim, 2023Serim}.
The steepness of this red component is comparable to the case of 4U 1626--67 \citep{1997Bildsten, 2023Serim}; however, the timescales within which they are observed are dissimilar.
The PDS continuum break occurs at $\omega_b\gtrsim 1/46$ days$^{-1}$ and evolves toward a flatter continuum at higher analysis frequencies (i.e., becomes a white noise component, S$_{r,2}=(6.76\pm0.16)\times10^{-19}$ Hz$^2$ s$^{-2}$ Hz$^{-1}$), implying uncorrelated torque fluctuations at shorter timescales.
When the regular frequency evolution model is substituted for the luminosity-dependent model, the power density estimate at the longest timescale is reduced by a factor of $>$100.
The steepness of the red noise component is also reduced to $\omega^{-0.91\pm0.38}$ but it does not completely vanish unlike the case of 2S 1417--624 \citep{2021Serim}.
It implies that either the luminosity-dependent model (at least through a simple power law relation) does not remove all of the red noise component of the PDS, or merely the standard disc component, which generally contributes to PDS spectra as $\omega^{-2}$ \citep{1997Bildsten, 2023Serim}, is subtracted from the PDS continuum.
At higher analysis frequencies ($\omega_b\approx 1/27$ days$^{-1}$), the PDS carries the same structure as the former case, with the white noise normalization S$_{r,2}=(5.25\pm0.17) \times 10^{-19}$ Hz$^2$ s$^{-2}$ Hz$^{-1}$.

\begin{figure}
\centering
	\includegraphics[width=\columnwidth]{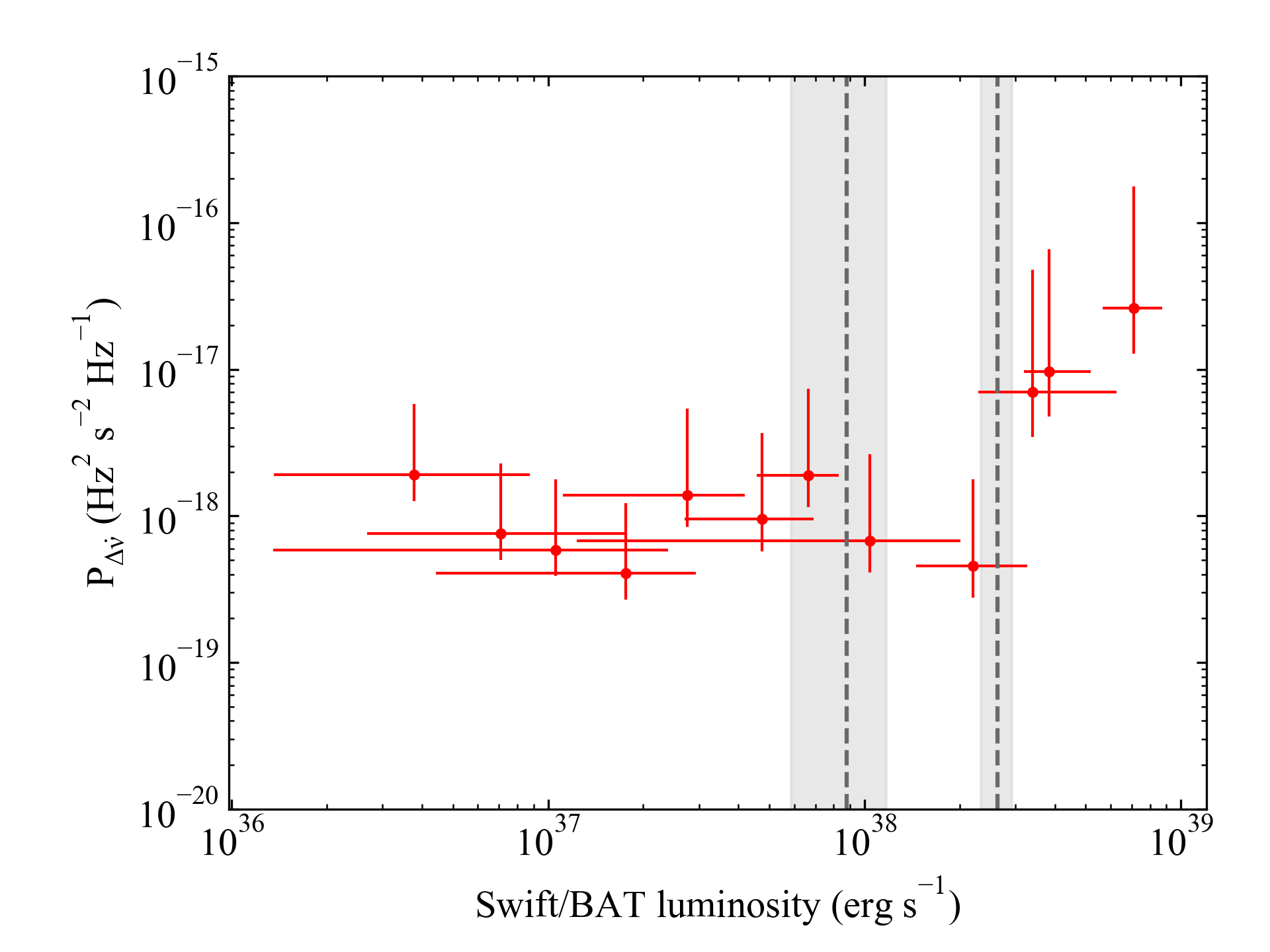}
    \caption{The distribution of noise strength estimates using the standard method on the timescale of 15 days as a function of luminosity. The dashed grey lines indicate the transitional luminosity levels $L_1$ (left) and $L_2$ (right), which are calculated for a distance of 5.2 kpc. The shaded regions reflect the uncertainties in the interpretation of the corresponding transitional luminosities in the literature. The measurements below $L_2$ are rebinned by a factor of 2 for visual purposes.}
    \label{fig:noisevslum}
\end{figure}

In order to understand the nature of the strong red noise component in the PDS of torque fluctuations, we further check the luminosity dependence of the timing noise strengths.
Hence, we split the frequency history into $\sim$15 days long segments and calculate the noise strengths for each of them using the standard method described above. Next, using the \emph{Swift}/BAT count rates, we calculate the luminosity range for each interval\footnote{See Section \ref{sec2} for the conversion of \emph{Swift}/BAT count rates to luminosity.}.
The distribution of the noise strength estimates as a function of luminosity is illustrated in Figure \ref{fig:noisevslum}.
The luminosity dependence of the noise strength estimates yields an intriguing distribution.
The noise strength amplitudes remain more or less constant up to the transitional luminosity level $L_2$ with a slight de-escalation between $L_1$ and $L_2$.
The $S_r$ values significantly rise (by a factor of $\sim$10) above $L_2$.
It indicates a possible change in the nature of torque fluctuations above $L_2$.

\section{Discussion and Conclusion}
\label{sec4}
We analyse the \emph{NICER}/XTI data set and enrich the spin frequency history of Swift J0243.6+6124 with new measurements.
The late-stage evolution of spin frequency indicates another torque reversal around MJD $\sim$58510, after which the source entered a new spin-down phase.
When the frequency evolution of Swift J0243.6+6124 is examined, the spin-down phases seem to occur systematically at luminosities below $\sim$$7\times10^{36}$ erg s$^{-1}$.
It has already been shown that the source pulsations were observable at luminosities down to 10$^{34}$--10$^{35}$ erg s$^{-1}$, implying that propeller stage is not yet attained at such low levels \citep{2018Tsygankov, 2020Doroshenko}, and therefore the spin-down phase is not associated with propeller regime.

 The pulse profile evolution of Swift J0243.6+6124 is very intriguing. At luminosities below $\sim$$7\times 10^{36} $ erg s$^{-1}$, the pulse profiles are single peaked. Between $\sim$ 7 $\times$ 10$^{36}$ erg s$^{-1}<L_x<L_1$, a secondary peak component emerges and gains strength with increasing luminosity; thus, the profiles become double peaked. Furthermore, when $L_x>L_1$, the pulse profiles become single-peaked again.
 The transformation of the pulse profiles around $L_x \sim 7 \times 10^{36}$ erg s$^{-1}$ indicates a new transition in the accretion geometry.
 According to \cite{2012Becker}, the critical X-ray luminosity ($L_1$) specifies the onset of the transition from fan beam to pencil beam; however, the transition does not immediately take place. There is an intermediate accretion regime $L_{coul}<L_x<L_1$ where the final phase of the deceleration of the accreted material is experienced through Coulomb braking in the plasma. In such a regime, a hybrid combination of both fan and pencil beam patterns is expected \citep{2012Becker,2000Blum}.   They specify a limiting luminosity below which Coulomb interactions are no longer effective enough to stop the accretion flow. This transition luminosity is given by:
\begin{equation}
 L_{coul} \simeq 1.17 \times 10^{37}  B_{12}^{-1/3} \Lambda_{0.1}^{-7/12} \tau_{20}^{7/12} M_{1.4}^{11/8} R_{10}^{13/24} \,\textrm{erg s}^{-1}
 \label{eqLcoul}
\end{equation}
where $B_{12}\equiv B/10^{12}$ G is the dipolar magnetic field strength of the pulsar, $\Lambda_{0.1} \equiv \Lambda/0.1$ is a dimensionless parameter accounting for various physical processes such as the possible role of plasma shielding, $\tau_{20} \equiv \tau/20$ is the Thomson optical depth, $M_{1.4} \equiv M/1.4$ $M_\odot$ is the pulsar mass, and $R_{10} \equiv R/10$ km is the pulsar radius. Below this luminosity, the accretion flow is suggested to be decelerated via gas-mediated shock near the stellar surface and the radiation from the polar caps fully transforms to a pencil beam pattern. In addition, according to this model, the pencil beam pattern should also persist at lower luminosity levels ($L_x<<L_{coul}$). Actually, such a single-peaked pulse profile was observed by \citet{2020Doroshenko} with an 80 ks \emph{NuSTAR} observation around the luminosity level of $\sim$3 $\times$ 10$^{34}$ erg s$^{-1}$.  If we consider the transition at $7 \times 10^{36}$ erg s$^{-1}$  as $L_{coul}$, neglecting the normalized dimensionless parameters of about unity and using typical neutron star parameters, then the magnetic field of the source can be estimated as $4.7\times10^{12}$ G. Furthermore, when the previously reported critical luminosity level \citep{2018WilsonHodge,2020Doroshenko},  $L_1$, of the onset of the transition from hybrid pattern to fan beam is updated for the same distance, it results in a magnetic field strength of $5.3\times 10^{12}$ G. Thus, magnetic field strength estimations obtained from both transitional levels become consistent at 5.2 kpc. Therefore, we suggest that the dipolar magnetic field strength of Swift J0243.6+6124 can be confined to a range of $\sim$$(4.7-5.3) \times 10^{12}$ G.

On the other hand, we also investigated the PDS of the spin frequency derivative fluctuations using the fairly sampled spin frequency data set which is improved with new measurements obtained from \emph{NICER}/XTI observations.
We extract two different PDSs using different models to describe regular rotational evolution.
The first one utilizes the standard polynomial-driven approach, and the second one makes use of a luminosity-dependent spin frequency evolution model.
Both PDSs exhibit bimodal behaviour in which the high analysis frequency ($\omega_b\sim 1/46$ days$^{-1}$ for the former, $\omega_b\sim1/27$ days$^{-1}$ for the latter case) noise components are flat while the low analysis frequency components carry red noise.
It should be noted that the observed break frequencies are rather close to the orbital period of the source ($\sim$27.7 d).
The white noise components in the PDS of the spin frequency derivative fluctuations are generally attributed to the uncorrelated torque fluctuations generated via wind accretion from the companion \citep{1972Boynton, 1985Deeter, 1997Bildsten, 2023Serim}.
Hence, the high analysis frequency white noise component of Swift J0243.6+6124 may hint at the accretion from the stellar wind of its companion, which is effective on timescales less than the orbital period of the source. Nevertheless, the long-term spin evolution and fluctuations are governed by the disc interactions.

In general, for the sources that are presumed to have an accretion disc, the red noise continuum with $\omega^{-2}$ dependence sets in at low timescales, which are possibly saturated at viscous timescales \citep{1997Bildsten, 2023Serim}.
Even though the number of studies is limited, the PDSs of the torque fluctuations of transient accreting sources demonstrate that steepness of red noise components also seem to occur as $\sim$$\omega^{-2}$ (e.g., SAX J2103.5+4545, \cite{2007Baykal}; 2S 1417--624, \cite{2021Serim}). 
Utilizing the standard PDS generation method, we find that the steepness of the red noise component of Swift J0243.6+6124 is significantly higher ($\sim$$\omega^{-3.36}$) when compared with other accreting sources \citep{2023Serim}.
Such a steep red noise component is only observed in ultra-compact binary system 4U 1626--67 \citep{1997Bildsten, 2023Serim} and in several magnetars \citep{2002Woods, 2019CerriSerim}; however, the timescales in which the component arise are different than the case of Swift J0243.6+6124.
To be more specific, the red noise component of the PDS of Swift J0243.6+6124 develops approximately on the orbital timescales, whereas the red noise in 4U 1626--67 is present on timescales longer than $\sim$1000 days \citep{1997Bildsten,2023Serim}.
In the case of SGR 1806--20 and SGR 1900+14, the red noise components are observed on timescales longer than $\sim$100 days and the onset timescale of the red noise components are attributed to a threshold for which these magnetars become burst active \citep{2002Woods}. Therefore, we believe that the strong red noise component observed in Swift J0243.6+6124 originates from different physical processes than the aforementioned cases.
To understand the nature of this component, we used the procedure described in \citep{2021Serim} where the rotational evolution is prescribed by a simple torque model.
In the case of 2S 1417--624 \citep{2021Serim}, this model almost completely eliminates the red noise component associated with disc accretion; however, for Swift J0243.6+6124, the results are slightly peculiar.
The steepness is reduced from $\sim\omega^{-3}$ to $\omega^{-1}$ but the red noise structure does not entirely vanish.
This situation may originate from different factors.
First, it is possible that the model used in Figure \ref{fig:torquelumfreq} provides an oversimplistic view for $\dot{\nu}$--$L_x$ correlation, thus more complex models (e.g., \cite{2022Karaferias}) are required to eliminate this component.
Secondly, if the $\omega^{-1}$ dependence has a physical origin, then it may indicate that $\omega^{-2}$ dependence observed for the disc component is subtracted.
Noting that the steepness and strength of the torque fluctuations are generally attributed to the nature of the magnetic field \citep{2002Woods, 2019CerriSerim}, it is possible that the remaining red noise component might be of magnetic origin.
Therefore, we further investigate the luminosity dependence of the noise strength estimations to inspect the nature of torque fluctuations at different levels (see Figure \ref{fig:noisevslum}).
We find that the noise strengths remain roughly constant up to the critical luminosity level $L_2$, above which the RPD accretion disc regime sets in \citep{2020Doroshenko}.
When the source luminosity exceeds $L_2$, the noise strength estimates suddenly increase by a factor of 10, which suggest a possible change in the nature of torque fluctuations above this level.
Moreover, it is recently shown that that the torque--luminosity relation of Swift J0243.6+6124 flattens at the RPD regime \citep{2022Karaferias, 2022Liu2}.
Hence, as the luminosity increases, the torque exertions become less efficient and more noisy, which may originate from the interactions with the quadruple components of the field \citep{2007Long}. In addition, the observed CRSF was evident only in certain pulse phases at the peak of the outburst, and it is attributed to the multipole component of the field \citep{2022Kong}.
Thus, the excess noise strength above the transitional level $L_2$ bolsters the idea that multipole components should play an important role in torque interactions at super-Eddington luminosity levels \citep{2020Doroshenko, 2022Kong}.

\section*{Acknowledgements}
The authors thank the referee for their valuable remarks that assisted in the improvement of this manuscript.
Authors acknowledge the support from TÜBİTAK (The Scientific and Technological Research Council of Turkey) through the research project MFAG 118F037.
The authors also thank Prof. Dr. S{\i}tk{\i} \c{C}a\u{g}da\c{s} \.{I}nam for his insightful comments.

\section*{Data Availability}
Whole X-ray data used in this study are publicly available. \emph{NICER}/XTI data can obtained through the High Energy Astrophysics Science Archive Research Center (\url{https://heasarc.gsfc.nasa.gov}). \emph{Swift}/BAT count rates can be acquired via Swift transient monitoring project (\url{https://swift.gsfc.nasa.gov/results/transients/weak/}) and \emph{Fermi}/GBM frequency measurements are available at GBM Accreting Pulsars Project website (\url{https://gammaray.msfc.nasa.gov/gbm/science/pulsars.html}).



\bibliographystyle{mnras}
\bibliography{article} 

\begin{thebibliography}{}
\makeatletter
\relax
\def\mn@urlcharsother{\let\do\@makeother \do\$\do\&\do\#\do\^\do\_\do\%\do\~}
\def\mn@doi{\begingroup\mn@urlcharsother \@ifnextchar [ {\mn@doi@}
  {\mn@doi@[]}}
\def\mn@doi@[#1]#2{\def\@tempa{#1}\ifx\@tempa\@empty \href
  {http://dx.doi.org/#2} {doi:#2}\else \href {http://dx.doi.org/#2} {#1}\fi
  \endgroup}
\def\mn@eprint#1#2{\mn@eprint@#1:#2::\@nil}
\def\mn@eprint@arXiv#1{\href {http://arxiv.org/abs/#1} {{\tt arXiv:#1}}}
\def\mn@eprint@dblp#1{\href {http://dblp.uni-trier.de/rec/bibtex/#1.xml}
  {dblp:#1}}
\def\mn@eprint@#1:#2:#3:#4\@nil{\def\@tempa {#1}\def\@tempb {#2}\def\@tempc
  {#3}\ifx \@tempc \@empty \let \@tempc \@tempb \let \@tempb \@tempa \fi \ifx
  \@tempb \@empty \def\@tempb {arXiv}\fi \@ifundefined
  {mn@eprint@\@tempb}{\@tempb:\@tempc}{\expandafter \expandafter \csname
  mn@eprint@\@tempb\endcsname \expandafter{\@tempc}}}

\bibitem[\protect\citeauthoryear{{Acuner}, {{\.I}nam}, {{\c{S}}ahiner},
  {Serim}, {Baykal}  \& {Swank}}{{Acuner} et~al.}{2014}]{2014Acuner}
{Acuner} Z.,  {{\.I}nam} S.~{\c{C}}.,  {{\c{S}}ahiner} {\c{S}}.,  {Serim}
  M.~M.,  {Baykal} A.,   {Swank} J.,  2014, \mn@doi [\mnras]
  {10.1093/mnras/stu1351}, \href
  {https://ui.adsabs.harvard.edu/abs/2014MNRAS.444..457A} {444, 457}

\bibitem[\protect\citeauthoryear{{Bailer-Jones}, {Rybizki}, {Fouesneau},
  {Mantelet}  \& {Andrae}}{{Bailer-Jones} et~al.}{2018}]{2018BailerJones}
{Bailer-Jones} C.~A.~L.,  {Rybizki} J.,  {Fouesneau} M.,  {Mantelet} G.,
  {Andrae} R.,  2018, \mn@doi [\aj] {10.3847/1538-3881/aacb21}, \href
  {https://ui.adsabs.harvard.edu/abs/2018AJ....156...58B} {156, 58}

\bibitem[\protect\citeauthoryear{{Bailer-Jones}, {Rybizki}, {Fouesneau},
  {Demleitner}  \& {Andrae}}{{Bailer-Jones} et~al.}{2021}]{2021Bailer}
{Bailer-Jones} C.~A.~L.,  {Rybizki} J.,  {Fouesneau} M.,  {Demleitner} M.,
  {Andrae} R.,  2021, \mn@doi [\aj] {10.3847/1538-3881/abd806}, \href
  {https://ui.adsabs.harvard.edu/abs/2021AJ....161..147B} {161, 147}

\bibitem[\protect\citeauthoryear{{Baykal}, {Inam}, {Stark}, {Heffner}, {Erkoca}
   \& {Swank}}{{Baykal} et~al.}{2007}]{2007Baykal}
{Baykal} A.,  {Inam} S.~{\c{C}}.,  {Stark} M.~J.,  {Heffner} C.~M.,  {Erkoca}
  A.~E.,   {Swank} J.~H.,  2007, \mn@doi [\mnras]
  {10.1111/j.1365-2966.2006.11231.x}, \href
  {https://ui.adsabs.harvard.edu/abs/2007MNRAS.374.1108B} {374, 1108}

\bibitem[\protect\citeauthoryear{{Becker} et~al.,}{{Becker}
  et~al.}{2012}]{2012Becker}
{Becker} P.~A.,  et~al., 2012, \mn@doi [\aap] {10.1051/0004-6361/201219065},
  \href {https://ui.adsabs.harvard.edu/abs/2012A&A...544A.123B} {544, A123}

\bibitem[\protect\citeauthoryear{{Bildsten} et~al.,}{{Bildsten}
  et~al.}{1997}]{1997Bildsten}
{Bildsten} L.,  et~al., 1997, \mn@doi [\apjs] {10.1086/313060}, \href
  {https://ui.adsabs.harvard.edu/abs/1997ApJS..113..367B} {113, 367}

\bibitem[\protect\citeauthoryear{{Blum} \& {Kraus}}{{Blum} \&
  {Kraus}}{2000}]{2000Blum}
{Blum} S.,  {Kraus} U.,  2000, \mn@doi [\apj] {10.1086/308308}, \href
  {https://ui.adsabs.harvard.edu/abs/2000ApJ...529..968B} {529, 968}

\bibitem[\protect\citeauthoryear{{Boynton}, {Groth}, {Hutchinson}, {Nanos},
  {Partridge}  \& {Wilkinson}}{{Boynton} et~al.}{1972}]{1972Boynton}
{Boynton} P.~E.,  {Groth} E.~J.,  {Hutchinson} D.~P.,  {Nanos} G.~P. J.,
  {Partridge} R.~B.,   {Wilkinson} D.~T.,  1972, \mn@doi [\apj]
  {10.1086/151550}, \href
  {https://ui.adsabs.harvard.edu/abs/1972ApJ...175..217B} {175, 217}

\bibitem[\protect\citeauthoryear{{Bykov}, {Gilfanov}, {Tsygankov}  \&
  {Filippova}}{{Bykov} et~al.}{2022}]{2022Bykov}
{Bykov} S.~D.,  {Gilfanov} M.~R.,  {Tsygankov} S.~S.,   {Filippova} E.~V.,
  2022, \mn@doi [\mnras] {10.1093/mnras/stac2239}, \href
  {https://ui.adsabs.harvard.edu/abs/2022MNRAS.516.1601B} {516, 1601}

\bibitem[\protect\citeauthoryear{{Cordes} \& {Downs}}{{Cordes} \&
  {Downs}}{1985}]{1985Cordes}
{Cordes} J.~M.,  {Downs} G.~S.,  1985, \mn@doi [\apjs] {10.1086/191076}, \href
  {https://ui.adsabs.harvard.edu/abs/1985ApJS...59..343C} {59, 343}

\bibitem[\protect\citeauthoryear{{Deeter}}{{Deeter}}{1984}]{1984Deeter}
{Deeter} J.~E.,  1984, \mn@doi [\apj] {10.1086/162122}, \href
  {https://ui.adsabs.harvard.edu/abs/1984ApJ...281..482D} {281, 482}

\bibitem[\protect\citeauthoryear{{Deeter} \& {Boynton}}{{Deeter} \&
  {Boynton}}{1985}]{1985Deeter}
{Deeter} J.~E.,  {Boynton} P.~E.,  1985, in {S. Hayakawa and F. Nagase} ed.,
  Inuyama Workshop on timing analysis of X-ray sources. p.~29

\bibitem[\protect\citeauthoryear{{Deeter}, {Boynton}, {Shibazaki}, {Hayakawa},
  {Nagase}  \& {Sato}}{{Deeter} et~al.}{1987}]{1987Deeter}
{Deeter} J.~E.,  {Boynton} P.~E.,  {Shibazaki} N.,  {Hayakawa} S.,  {Nagase}
  F.,   {Sato} N.,  1987, \mn@doi [\aj] {10.1086/114373}, \href
  {https://ui.adsabs.harvard.edu/abs/1987AJ.....93..877D} {93, 877}

\bibitem[\protect\citeauthoryear{{Doroshenko}, {Tsygankov}  \&
  {Santangelo}}{{Doroshenko} et~al.}{2018}]{2018Doroshenko}
{Doroshenko} V.,  {Tsygankov} S.,   {Santangelo} A.,  2018, \mn@doi [\aap]
  {10.1051/0004-6361/201732208}, \href
  {https://ui.adsabs.harvard.edu/abs/2018A&A...613A..19D} {613, A19}

\bibitem[\protect\citeauthoryear{{Doroshenko} et~al.,}{{Doroshenko}
  et~al.}{2020}]{2020Doroshenko}
{Doroshenko} V.,  et~al., 2020, \mn@doi [\mnras] {10.1093/mnras/stz2879}, \href
  {https://ui.adsabs.harvard.edu/abs/2020MNRAS.491.1857D} {491, 1857}

\bibitem[\protect\citeauthoryear{{Gendreau} et~al.,}{{Gendreau}
  et~al.}{2016}]{2016Gendreau}
{Gendreau} K.~C.,  et~al., 2016, in {den Herder} J.-W.~A.,  {Takahashi} T.,
  {Bautz} M.,  eds,  Society of Photo-Optical Instrumentation Engineers (SPIE)
  Conference Series Vol. 9905, Space Telescopes and Instrumentation 2016:
  Ultraviolet to Gamma Ray. p. 99051H, \mn@doi{10.1117/12.2231304}

\bibitem[\protect\citeauthoryear{{Jaisawal}, {Naik}  \& {Chenevez}}{{Jaisawal}
  et~al.}{2018}]{2018Jaisawal}
{Jaisawal} G.~K.,  {Naik} S.,   {Chenevez} J.,  2018, \mn@doi [\mnras]
  {10.1093/mnras/stx3082}, \href
  {https://ui.adsabs.harvard.edu/abs/2018MNRAS.474.4432J} {474, 4432}

\bibitem[\protect\citeauthoryear{{Jaisawal} et~al.,}{{Jaisawal}
  et~al.}{2019}]{2019Jaisawal}
{Jaisawal} G.~K.,  et~al., 2019, \mn@doi [\apj] {10.3847/1538-4357/ab4595},
  \href {https://ui.adsabs.harvard.edu/abs/2019ApJ...885...18J} {885, 18}

\bibitem[\protect\citeauthoryear{{Karaferias}, {Vasilopoulos}, {Petropoulou},
  {Jenke}, {Wilson-Hodge}  \& {Malacaria}}{{Karaferias}
  et~al.}{2022}]{2022Karaferias}
{Karaferias} A.~S.,  {Vasilopoulos} G.,  {Petropoulou} M.,  {Jenke} P.~A.,
  {Wilson-Hodge} C.~A.,   {Malacaria} C.,  2022, \mn@doi [arXiv e-prints]
  {10.48550/arXiv.2211.16079}, \href
  {https://ui.adsabs.harvard.edu/abs/2022arXiv221116079K} {p. arXiv:2211.16079}

\bibitem[\protect\citeauthoryear{{Kennea}, {Lien}, {Krimm}, {Cenko}  \&
  {Siegel}}{{Kennea} et~al.}{2017}]{2017Kennea}
{Kennea} J.~A.,  {Lien} A.~Y.,  {Krimm} H.~A.,  {Cenko} S.~B.,   {Siegel}
  M.~H.,  2017, The Astronomer's Telegram, \href
  {https://ui.adsabs.harvard.edu/abs/2017ATel10809....1K} {10809, 1}

\bibitem[\protect\citeauthoryear{{Kong} et~al.,}{{Kong}
  et~al.}{2020}]{2020Kong}
{Kong} L.~D.,  et~al., 2020, \mn@doi [\apj] {10.3847/1538-4357/abb241}, \href
  {https://ui.adsabs.harvard.edu/abs/2020ApJ...902...18K} {902, 18}

\bibitem[\protect\citeauthoryear{{Kong} et~al.,}{{Kong}
  et~al.}{2022}]{2022Kong}
{Kong} L.-D.,  et~al., 2022, \mn@doi [\apjl] {10.3847/2041-8213/ac7711}, \href
  {https://ui.adsabs.harvard.edu/abs/2022ApJ...933L...3K} {933, L3}

\bibitem[\protect\citeauthoryear{{Kouroubatzakis}, {Reig}, {Andrews}  \&
  {)}}{{Kouroubatzakis} et~al.}{2017}]{2017Kouroubatzakis}
{Kouroubatzakis} K.,  {Reig} P.,  {Andrews} J.,   {)} A.~Z.,  2017, The
  Astronomer's Telegram, \href
  {https://ui.adsabs.harvard.edu/abs/2017ATel10822....1K} {10822, 1}

\bibitem[\protect\citeauthoryear{{Krimm} et~al.,}{{Krimm}
  et~al.}{2013}]{2013Krimm}
{Krimm} H.~A.,  et~al., 2013, \mn@doi [\apjs] {10.1088/0067-0049/209/1/14},
  \href {https://ui.adsabs.harvard.edu/abs/2013ApJS..209...14K} {209, 14}

\bibitem[\protect\citeauthoryear{{Liu}, {Vasilopoulos}, {Ge}, {Ji}, {Weng},
  {Zhang}  \& {Hou}}{{Liu} et~al.}{2022a}]{2022Liu2}
{Liu} J.,  {Vasilopoulos} G.,  {Ge} M.,  {Ji} L.,  {Weng} S.-S.,  {Zhang}
  S.-N.,   {Hou} X.,  2022a, \mn@doi [\mnras] {10.1093/mnras/stac2746}, \href
  {https://ui.adsabs.harvard.edu/abs/2022MNRAS.517.3354L} {517, 3354}

\bibitem[\protect\citeauthoryear{{Liu} et~al.,}{{Liu} et~al.}{2022b}]{2022Liu}
{Liu} W.,  et~al., 2022b, \mn@doi [\aap] {10.1051/0004-6361/202243878}, \href
  {https://ui.adsabs.harvard.edu/abs/2022A&A...666A.110L} {666, A110}

\bibitem[\protect\citeauthoryear{{Long}, {Romanova}  \& {Lovelace}}{{Long}
  et~al.}{2007}]{2007Long}
{Long} M.,  {Romanova} M.~M.,   {Lovelace} R.~V.~E.,  2007, \mn@doi [\mnras]
  {10.1111/j.1365-2966.2006.11192.x}, \href
  {https://ui.adsabs.harvard.edu/abs/2007MNRAS.374..436L} {374, 436}

\bibitem[\protect\citeauthoryear{{Malacaria}, {Jenke}, {Roberts},
  {Wilson-Hodge}, {Cleveland}, {Mailyan}  \& {GBM Accreting Pulsars Program
  Team}}{{Malacaria} et~al.}{2020}]{2020Malacaria}
{Malacaria} C.,  {Jenke} P.,  {Roberts} O.~J.,  {Wilson-Hodge} C.~A.,
  {Cleveland} W.~H.,  {Mailyan} B.,   {GBM Accreting Pulsars Program Team}
  2020, \mn@doi [\apj] {10.3847/1538-4357/ab855c}, \href
  {https://ui.adsabs.harvard.edu/abs/2020ApJ...896...90M} {896, 90}

\bibitem[\protect\citeauthoryear{{Reig}, {Fabregat}  \&
  {Alfonso-Garz{\'o}n}}{{Reig} et~al.}{2020}]{2020Reig}
{Reig} P.,  {Fabregat} J.,   {Alfonso-Garz{\'o}n} J.,  2020, \mn@doi [\aap]
  {10.1051/0004-6361/202038333}, \href
  {https://ui.adsabs.harvard.edu/abs/2020A&A...640A..35R} {640, A35}

\bibitem[\protect\citeauthoryear{{Serim}, {{\"O}z{\"u}do{\u{g}}ru},
  {D{\"o}nmez}, {{\c{S}}ahiner}, {Serim}, {Baykal}  \& {{\.I}nam}}{{Serim}
  et~al.}{2022}]{2021Serim}
{Serim} M.~M.,  {{\"O}z{\"u}do{\u{g}}ru} {\"O}.~C.,  {D{\"o}nmez} {\c{C}}.~K.,
  {{\c{S}}ahiner} {\c{S}}.,  {Serim} D.,  {Baykal} A.,   {{\.I}nam}
  S.~{\c{C}}.,  2022, \mn@doi [\mnras] {10.1093/mnras/stab3547}, \href
  {https://ui.adsabs.harvard.edu/abs/2022MNRAS.510.1438S} {510, 1438}

\bibitem[\protect\citeauthoryear{{Serim}, {Serim}  \& {Baykal}}{{Serim}
  et~al.}{2023}]{2023Serim}
{Serim} D.,  {Serim} M.~M.,   {Baykal} A.,  2023, \mn@doi [\mnras]
  {10.1093/mnras/stac3076}, \href
  {https://ui.adsabs.harvard.edu/abs/2023MNRAS.518....1S} {518, 1}

\bibitem[\protect\citeauthoryear{{Sugizaki}, {Oeda}, {Kawai}, {Mihara},
  {Makishima}  \& {Nakajima}}{{Sugizaki} et~al.}{2020}]{2020Sugizaki}
{Sugizaki} M.,  {Oeda} M.,  {Kawai} N.,  {Mihara} T.,  {Makishima} K.,
  {Nakajima} M.,  2020, \mn@doi [\apj] {10.3847/1538-4357/ab93c7}, \href
  {https://ui.adsabs.harvard.edu/abs/2020ApJ...896..124S} {896, 124}

\bibitem[\protect\citeauthoryear{{Tao}, {Feng}, {Zhang}, {Bu}, {Zhang}, {Qu}
  \& {Zhang}}{{Tao} et~al.}{2019}]{2019Tao}
{Tao} L.,  {Feng} H.,  {Zhang} S.,  {Bu} Q.,  {Zhang} S.,  {Qu} J.,   {Zhang}
  Y.,  2019, \mn@doi [\apj] {10.3847/1538-4357/ab0211}, \href
  {https://ui.adsabs.harvard.edu/abs/2019ApJ...873...19T} {873, 19}

\bibitem[\protect\citeauthoryear{{Tsygankov}, {Doroshenko}, {Mushtukov},
  {Lutovinov}  \& {Poutanen}}{{Tsygankov} et~al.}{2018}]{2018Tsygankov}
{Tsygankov} S.~S.,  {Doroshenko} V.,  {Mushtukov} A.~A.,  {Lutovinov} A.~A.,
  {Poutanen} J.,  2018, \mn@doi [\mnras] {10.1093/mnrasl/sly116}, \href
  {https://ui.adsabs.harvard.edu/abs/2018MNRAS.479L.134T} {479, L134}

\bibitem[\protect\citeauthoryear{{Wang} et~al.,}{{Wang}
  et~al.}{2020}]{2020Wang}
{Wang} P.~J.,  et~al., 2020, \mn@doi [\mnras] {10.1093/mnras/staa2448}, \href
  {https://ui.adsabs.harvard.edu/abs/2020MNRAS.497.5498W} {497, 5498}

\bibitem[\protect\citeauthoryear{{Wilson-Hodge} et~al.,}{{Wilson-Hodge}
  et~al.}{2018}]{2018WilsonHodge}
{Wilson-Hodge} C.~A.,  et~al., 2018, \mn@doi [\apj] {10.3847/1538-4357/aace60},
  \href {https://ui.adsabs.harvard.edu/abs/2018ApJ...863....9W} {863, 9}

\bibitem[\protect\citeauthoryear{{Woods}, {Kouveliotou},
  {G{\"o}{\v{g}}{\"u}{\c{s}}}, {Finger}, {Swank}, {Markwardt}, {Hurley}  \&
  {van der Klis}}{{Woods} et~al.}{2002}]{2002Woods}
{Woods} P.~M.,  {Kouveliotou} C.,  {G{\"o}{\v{g}}{\"u}{\c{s}}} E.,  {Finger}
  M.~H.,  {Swank} J.,  {Markwardt} C.~B.,  {Hurley} K.,   {van der Klis} M.,
  2002, \mn@doi [\apj] {10.1086/341536}, \href
  {https://ui.adsabs.harvard.edu/abs/2002ApJ...576..381W} {576, 381}

\bibitem[\protect\citeauthoryear{{{\c{C}}erri-Serim}, {Serim}, {{\c{S}}ahiner},
  {Inam}  \& {Baykal}}{{{\c{C}}erri-Serim} et~al.}{2019}]{2019CerriSerim}
{{\c{C}}erri-Serim} D.,  {Serim} M.~M.,  {{\c{S}}ahiner} {\c{S}}.,  {Inam}
  S.~{\c{C}}.,   {Baykal} A.,  2019, \mn@doi [\mnras] {10.1093/mnras/sty3213},
  \href {https://ui.adsabs.harvard.edu/abs/2019MNRAS.485....2C} {485, 2}

\bibitem[\protect\citeauthoryear{van~den {Eijnden}, {Degenaar}, {Russell},
  {Wijnands}, {Miller-Jones}, {Sivakoff}  \& {Hern{\'a}ndez
  Santisteban}}{van~den {Eijnden} et~al.}{2018}]{2018VanDenEijnden}
van~den {Eijnden} J.,  {Degenaar} N.,  {Russell} T.~D.,  {Wijnands} R.,
  {Miller-Jones} J.~C.~A.,  {Sivakoff} G.~R.,   {Hern{\'a}ndez Santisteban}
  J.~V.,  2018, \mn@doi [\nat] {10.1038/s41586-018-0524-1}, \href
  {https://ui.adsabs.harvard.edu/abs/2018Natur.562..233V} {562, 233}

\makeatother
\end{thebibliography}



\bsp	
\label{lastpage}
\end{document}